\documentclass[notitlepage]{article}
\usepackage{amssymb}

\newtheorem{theorem}{Theorem}[section]

\newtheorem{lemma}{Lemma}[section]

\newcommand{\R}{\mathbb{ R}}

\newcommand{\pr}{\mathrm{ pr}}
\newcommand{\diag}{\mathrm{ diag}}

\begin{document}
\pagestyle{plain}

\title{Some multidimensional integrable cases of nonholonomic
rigid body dynamics \footnote{Journal Ref.: Regular $\&$ Chaotic
Dynamics, 2003, {\bf 8} (1), 125-132.}}

\author{Bo\v zidar Jovanovi\' c \\
\small Mathematical Institute, SANU\\
\small Kneza Mihaila 35, 11000 Belgrade, Serbia\\
\small{\tt e-mail: bozaj$@$mi.sanu.ac.yu}}
\date{}
\maketitle

\begin{abstract}
In this paper we study the dynamics of  the
constrained $n$--dimensional rigid body (the Suslov problem).
We give a review of known integrable cases in three dimensions
and present their higher dimensional generalizations.
\end{abstract}

\section{The Suslov problem}

The equations of  nonholonomic systems are not Hamiltonian. They are
Hamiltonian with respect to the almost-Poisson brackets \cite{CLM}.
This is the reason that the integration theory of constrained mechanical
systems is less developed then for unconstrained ones.
However, in some solvable nonholonomic systems with an invariant measure,
the phase space is foliated by invariant tori, placing these  systems together with integrable
Hamiltonian systems (see \cite{Ch, AKN, KZ, VV, Ko1, FK}).

In this paper we are interested in the integrability of  multidimensional
generalizations of the Suslov nonholonomic rigid body problem.

Consider an $n$--dimensional rigid body motion around the
fixed point $O=(0,0,\dots,0)$ in $\R^n$. The configuration space
is the Lie group $SO(n)$. The matrix $g\in SO(n)$ maps the orthonormal frame $E_1,\dots,E_n$
fixed in the space to the
frame $e_1=g\cdot E_1,\dots,e_n=g\cdot E_n$ fixed in the body,
$e_1=(1,0,\dots,0)^t,\dots,e_n=(0,\dots,0,1)^t$.  For
a motion $g(t)\in SO(n)$, the {\it angular velocity} and {\it momentum} (in body coordinates) are
$\Omega(t)=g^{-1}\cdot g(t)$ and $M=J(\Omega)$ respectively. $J: so(n) \to so(n)^*$
is the {\it inertia tensor} and has the form:
$M=J(\Omega)=I\Omega+\Omega I$, where $I$ is symmetric $n\times n$ matrix called {\it mass tensor} (see \cite{FK}).
Here we identified $so(n)$ and $so(n)^*$ by the Killing scalar product.

Suppose that additional left--invariant constraints
\begin{equation}
\langle a^i,\Omega\rangle=0, \quad i=1,\dots,r \label{2.1}
\end{equation}
and potential force $V(\Gamma)$, where $\Gamma$ is a unit vector
fixed in the space, are given.
Let $D=\{\Omega\in so(n), \,\langle a^i,\Omega\rangle=0, \, i=1,\dots,r\}$ and $\mathcal D$ be the
corresponding left--invariant distribution of $TSO(n)$. The distribution
$\mathcal D$ is nonintegrable
(i.e., constraints are nonholonomic) if and only if $D$ is not a  subalgebra of $so(n)$.

The equations of motion of a rigid body subjected to the constraints (\ref{2.1}) and the potential force $V(\Gamma)$ are
\begin{equation}
\dot M=[M,\Omega]+
\frac{\partial V}{\partial \Gamma}\wedge\Gamma +\sum_{i=1}^r \lambda_i a^i,
\quad\dot \Gamma=-\Omega\cdot\Gamma,
\label{2.2}
\end{equation}
where Lagrange multipliers $\lambda_i$ are determined from the constraints (\ref{2.1}).
To describe a motion $g(t)$
on $SO(n)$ one should solve an additional equation $\Omega=g^{-1}\cdot g$.

For $n=3$, $r=1$, the equations (\ref{2.1}), (\ref{2.2})
are equations of classical {\it Suslov problem} (see \cite{Su, Ko1, AKN}).
Therefore, the above system can be considered as a natural multidimensional
generalization of the Suslov problem.

If $D$ is an eigenspace of $J$ then the equations (\ref{2.2}) take
the form
\begin{equation}
J(\dot\Omega)=\pr_D [J(\Omega),\Omega]+\pr_{D}(\frac{\partial V}{\partial \Gamma}\wedge\Gamma),
\quad
\dot \Gamma=-\Omega\Gamma, \quad \Omega\in D
\label{2.3}
\end{equation}

Contrary to the general case (\ref{2.2}), system (\ref{2.3})
preserves the standard measure $d\Omega d\Gamma$ on $ \mathcal
M=D\{\Omega\}\times S^{n-1}\{\Gamma\}, \quad
S^{n-1}=\{\Gamma\in\R^n, \, \langle\Gamma,\Gamma\rangle=1\}. $ The
energy $E= \frac12 \langle J\Omega ,\Omega \rangle  +V(\Gamma)$ is
always an integral of (\ref{2.2}), (\ref{2.3}). If, besides the
Hamiltonian function, we have $\dim D+n-4$ additional independent
integrals, by the Jacobi theorem the solutions of (\ref{2.3}) can
be found by quadratures. Under the compactness conditions, the
phase space $\mathcal M$ is foliated by invariant tori with
quasi-periodic dynamic (after appropriate change of time),
similarly as in the Liouville theorem \cite{Ko1, AKN}.

Therefore,
it is natural to call the system {\it completely integrable} if it can be integrated by the Jacobi theorem; or more
generally (see \cite{VV}): if the trajectories of the system belong to invariant tori
with the dynamic of the form
\begin{equation}
\dot \varphi_1=\frac{\omega_1}{\Phi(\varphi_1,\dots,\varphi_k)},\dots,
\dot \varphi_k=\frac{\omega_k}{\Phi(\varphi_1,\dots,\varphi_k)}.
\label{d1.4}
\end{equation}
The flow of (\ref{d1.4}) in the new time $d\tau=\Phi^{-1} dt$ is quasi-periodic: $d\varphi_i/d\tau=\omega_i$. Moreover,
for almost all frequencies $\omega_1,\dots,\omega_k$, by smooth change of variables $\theta_i=\theta_i(\varphi)$ equations
(\ref{d1.4}) can be reduced to the form $\dot \theta_i=\Omega_i=\omega_i/\Pi$, $i=1,\dots,k$, where $\Pi$ denotes the
total measure of $\mathbb{T}^k$ (see \cite{AKN}).

Note that the integrability of equations $(\ref{2.2})$ without the
constraints (for example the multidimensional Euler \cite{Man},
Lagrange \cite{Be} and Klebsh \cite{Pe} cases) implies {\it
non-commutative integrability} of unconstrained  system. The phase
space $T^*SO(n)$ is foliated on $d \le \dim so(n)$ dimensional
invariant isotropic tori with quasi-periodic dynamics (see \cite{MF, AKN, BJ}).
In the nonholonomic case there is no Poisson structure.
In order to precisely describe dynamic on the whole
phase space $\mathcal D$  we have to solve kinematic equation
$g^{-1}(t)\cdot \dot g(t)=\Omega(t)$. This problem, for the
Fedorov--Kozlov integrable case, is studied by Zenkov and Bloch
\cite{ZB}.

In the next section we give a review of the known integrable cases in three dimension. Then we present their
natural higher dimensional generalizations. It appears that these systems are also completely integrable (section 3).

\section{Three--dimensional integrable cases}

For $n=3$ we can identify the Lie algebra $(so(3),[\cdot,\cdot])$ with $(\R^3,\times)$ via
$\Omega=(\Omega_1,\Omega_2,\Omega_3)=(-\Omega_{23},\Omega_{13},-\Omega_{12})$. Then
the equations of the motion (\ref{2.2}) become:
\begin{equation}
J\dot \Omega =J\Omega \times \Omega +\Gamma \times \frac {\partial V}{\partial \Gamma } +\lambda a,
\quad\dot \Gamma = \Gamma \times \Omega, \quad \langle a, \Omega \rangle =0,  \label{3.1}
\end{equation}

Without loss of generality we can take $a=(0,0,1)$, i.e., the constraint is $\Omega_3=-\Omega_{12}=0$. Then
$D=\R^2\{\Omega_1,\Omega_2\}$. Geometrically, this means that only infinitesimal rotations in the planes
$e_1\wedge e_3$ and $e_2 \wedge e_3$ are allowed.

If $V=0$, then the equations (\ref{3.1}) form a closed system in variables $\Omega_1,\Omega_2$. If $a$ is not an eigen
vector of $J$ then there is an asymptotic line $l$ in $D$.
The angular velocity is moving along the energy ellipse
$\mathcal E=\{E(\Omega)=h\}$ asymptotically to
intersections points
$w_{-}$ and $w_{+}$ of $\mathcal E$ with $l$
(see Suslov \cite{Su}).
That system, obviously,
has not an invariant measure. When $a$ is an
eigenvector of $J$, then system (\ref{3.1}) preserves the measure $d\Omega d\Gamma$ and for $V=0$ the solutions
are very simple $\Omega=const$.

Thus, further we assume that $a$ is an eigenvector of the inertia tensor $J$. We can take $J=\diag(J_1,J_2,J_3)$.
For the integrability, we need one additional integral, independent of the Hamiltonian function.

The known integrable cases are:

\begin{itemize}

\item The Lagrange case (see Kozlov \cite{Ko1}), when $J_1=J_2$ and $V=B_3\Gamma_3$. The integral is
$F=\langle J \Omega, \Gamma \rangle $.

\item  The Kharlamova case \cite{KZ, AKN} where $V(\Gamma)=B_1\Gamma_1+B_2\Gamma_2$. Then
$F=J_1\Omega_1B_1+J_2\Omega_2 B_2$.

\item The Klebsh--Tisserand case (Kozlov \cite{Ko1, AKN}) with
$V(\Gamma )=\frac {\epsilon }2\langle J\Gamma, \Gamma \rangle $ and
$F=\frac 12 \langle J\Omega ,J\Omega\rangle -\frac 12 \langle A\Gamma , \Gamma \rangle $, where $A=\epsilon J^{-1}\det J$.

\end{itemize}

Let us note that in the first case the Lagrange multiplier is
$\lambda =0$, and this is actually an invariant
subsystem of the usual Lagrange rigid body system.  Also, Ziglin proved that if $V(\Gamma)=B_3\Gamma_3$ than system
(\ref{3.1}) has a complementary meromorphic first integral in the complexified phase space
$\mathcal M_\mathbb C=\mathbb{C}^2\{\Omega_1,\Omega_2\}\times S^2_{\mathbb{C}}\{\Gamma\}$ only in the Lagrange case
$J_1=J_2$ \cite{Zi}.

The Kharlamova and Klebsh-Tisserand cases
can be uniquely described and generalized in the following way (see \cite{DGJ}). If $J_1 \ne J_2$ then the  equations
(\ref{3.1}) of the Suslov problem are integrable for the potentials:
$V(\Gamma) =v_1(\Gamma_1,\Gamma_2^2+\Gamma_3^2)+v_2(\Gamma_2,\Gamma_1^2+ \Gamma_3^2)$ where $v_1$ and $v_2$ are arbitrary functions of two variables. The corresponding third integral is:
$F=\frac 12 \langle J\Omega , J\Omega \rangle +J_2 v_1(\Gamma_1,\Gamma_2^2+\Gamma_3^2)+J_1 v_2(\Gamma_2,\Gamma_1^2+\Gamma_3^2).$

In addition, there are integrable cases of the Suslov problem, which include the gyroscopic force with the momentum
$\epsilon\Gamma\times\Omega$ added to the right hand side of (\ref{3.1}).
Since the gyroscopic force is conservative,
the Hamiltonian remains to be the first integral.
As well as the usual Suslov problem, the system with the gyroscopic force is integrable
for the potentials $V=0$, $V(\Gamma )=B_1\Gamma_1+B_2\Gamma_2$ and
$V(\Gamma)=\frac 12(a_1\gamma_1^2+a_2\gamma_2^2+a_3\gamma_3^2)$ (see \cite{DGJ}).

\section{Multidimensional integrable cases}

Without potential force, the equations (\ref{2.1}), (\ref{2.2}) represent a closed system in variable
$\Omega$ ({\it Euler--Poincare--Suslov} equations). We can consider
Euler--Poincare--Suslov equations on other Lie algebras
as well. They are reduced equations of nonholonomic systems on Lie groups with left--invariant kinetic energy and
left--invariant constraints (see \cite{Ko1, Koi}). Integrable
examples can be found in \cite{FK, Jo1, Jo2}.

Since in the three--dimensional case, only infinitesimal rotations in the planes
$e_1\wedge e_3$ and $e_2 \wedge e_3$ are
allowed, Fedorov and Kozlov suggested that multidimensional analogue
of Suslov's conditions can be define in the
following way: only infinitesimal rotations in the planes $e_1\wedge e_n,\dots,e_{n-1} \wedge e_n$, i.e., in the planes
containing the vector $e_n$ are allowed \cite{FK}.
Therefore, we have the following constraints imposed to the components
of the angular velocity matrix:
\begin{equation}
\Omega_{ij}=0, \quad i,j \le n-1.
\label{4.1}
\end{equation}
Then
\begin{equation}
\mathcal M=\R^{n-1}\{\Omega_{1n},\Omega_{2n},\dots,\Omega_{n-1,n}\} \times S^{n-1}\{\Gamma\}.
\label{4.1a}
\end{equation}

Really, for $V=0$, the system has the same behavior as
three--dimensional system. If $D$ is an eigen space of the inertia
operator $J$ then the solutions are simply $\Omega=const$. In
general, all trajectories lying on the energy ellipsoid $\mathcal
E=\frac12\langle J\Omega,\Omega\rangle=h$ are double asymptotic:
they tend to the diametrically opposed points $w_+$ and $w_-$ as
$t\to\pm\infty$ (see \cite{FK}).

With the choice of constraints (\ref{4.1}) we shall see that
three--dimensional integrable cases with invariant measure have
their multidimensional analogues.

Suppose that the mass tensor is diagonal $I=\diag(I_1,I_2,\dots,I_n)$. Then $D$ is an invariant subspace of $J$.
The orthogonal complement of $D$ in $so(n)$ is $so(n-1)$.
Since
$(so(n),so(n-1))$ is a symmetric pair ($[so(n-1),so(n-1)]\subset so(n-1)$, $[so(n-1),D]\subset D$,
$[D,D]\subset so(n-1)$) the equations (\ref{2.3}) take the following simple form:
$$
J(\dot\Omega)=\pr_{D}(\frac{\partial V}{\partial \Gamma}\wedge\Gamma),\quad
 \dot \Gamma=-\Omega\Gamma, \quad \Omega\in D,
$$
or coordinately:
\begin{eqnarray}
&&(I_i+I_n)\dot\Omega_{in}=\frac{\partial V}{\partial \Gamma_i}\Gamma_n-\Gamma_i\frac{\partial V}{\partial \Gamma_n},
 \nonumber\\
&&\dot \Gamma_i =- \Gamma_n\Omega_{in}, \quad i=1,2,\dots,n-1,  \nonumber \\
&&\dot \Gamma_n = \Gamma_1\Omega_{1n}+\Gamma_2\Omega_{2n}+ \dots+\Gamma_{n-1}\Omega_{n-1,n}.
\label{4.3}
\end{eqnarray}

\subsection{The Lagrange case}

Let the mass tensor be of the form $I=\diag(I_1,\dots,I_1,I_n)$,
the rigid body be placed in a
homogeneous gravitational field and the center of mass be
on the axes of the dynamical symmetry
$e_n$. Then the potential is of the
form $V(\Gamma)=B_n\Gamma_n$ and system (\ref{4.3}) can be seen as an invariant subsystem of
the unconstrained
$n$--dimensional  rigid body motion:
\begin{eqnarray}
&&(I_i+I_j)\dot\Omega_{ij}= (I_i-I_j)\sum_k \Omega_{ik}\Omega_{kj}=0, \quad i,j\le n-1,\nonumber \\
&&(I_1+I_n)\dot\Omega_{in}= (I_1-I_n)\sum_k \Omega_{1k}\Omega_{kn} -B_n\Gamma_n, \nonumber \\
&&\dot \Gamma_i =-\sum_k \Omega_{ik}\Gamma_k, \quad i=1,\dots,n.
\label{4.4}
\end{eqnarray}
The Hamiltonian equations (\ref{4.4}) represent the natural generalization of the
three--dimensional Lagrange rigid body system. Their
integrability  is proved by Beljaev \cite{Be}.
However, this does not give us directly the integrability  on the invariant
subspace $M_{ij}=\Omega_{ij}=0$, $i,j\le n-1$.
In order to prove the integrability of the Suslov problem we shall use
the symplectic reduction,
instead of verifying independence of Beljaev's integrals on the invariant
subspace.

Consider the Lagrange rigid body system on the whole phase space
$T^*SO(n)$. It is invariant with respect to the {\it right}
Hamiltonian action of the group $SO(n-1)$ (rotations which fix
symmetry axes $e_n$). The corresponding moment map $\Phi:
T^*SO(n)\to so(n)^*$ is given by functions $M_{ij}$, $i,j\le n-1$
considered as the {\it left} invariant functions on $T^*SO(n)$.
One can easily prove:

\begin{lemma}
The reduced system on
$\Phi^{-1}(0)/SO(n-1)=T^*(SO(n)/SO(n-1))=T^*S^{n-1}$ is the spherical pendulum.
\end{lemma}

The spherical pendulum is a completely integrable system. Therefore,
the invariant submanifold $\Phi^{-1}(0)\subset T^*SO(n)$ of the Lagrange system,
as well as the rest of the phase space,
is almost everywhere foliated by invariant tori with quasi-periodic dynamics.
(More about the
relationships between integrability and reductions one can find in \cite{Zu, Jo3}.)
The Suslov problem on $\mathcal D\subset TSO(n)$, after Legendre transformation,
coincides with the Lagrange system on $\Phi^{-1}(0)$. Thus we obtain:

\begin{theorem}
The Suslov nonholonomic rigid body problem with constraints (\ref{4.1}), the mass tensor
of the form $I=\diag(I_1,\dots,I_1,I_n)$
and the center of the mass on the axes of the dynamical symmetry $e_n$,
in a presence of a
homogeneous gravitational field
(the multidimensional Lagrange case) is completely integrable.
\end{theorem}

Note that the unconstrained system on $T^*SO(n)$ is invariant
with respect to the {\it left} Hamiltonian action of the
group $SO(n-1)$ (rotations which fix $\Gamma$), for any inertia
tensor $J$ and potential $V(\Gamma)$. In this sense, the Lagrange
system admits two different $SO(n-1)$ reductions.
For $n=3$ this corresponds to the fact that
two Eulerian angles, the angle of precession $\psi$ and the angle of pure
rotations $\varphi$, are cyclic variables.
Elimination of the variable $\varphi$ with the zero value of the conjugate
momentum $p_\varphi=0$ gives the spherical pendulum on $S^2$
(for example, see \cite{BM}).

\subsection{The Kharlamova case}

The natural multidimensional generalization of the Kharlamova case can be defined as the constrained rigid
body motion (\ref{4.3}) in a gravitational field with the center of mass laying in  the plane spanned by
$e_1,e_2,\dots,e_{n-1}$. Then the potential force is of the form $V=B_1\Gamma_1+B_2\Gamma_2+\dots+B_{n-1}\Gamma_{n-1}$ and
equations (\ref{4.3}) become:
\begin{eqnarray}
&&(I_i+I_n)\dot\Omega_{in}=B_i\Gamma_n, \nonumber\\
&&\dot \Gamma_i=- \Gamma_n\Omega_{in}, \quad i=1,2,\dots,n-1,  \nonumber\\
&& \dot \Gamma_n= \Gamma_1\Omega_{1n}+\Gamma_2\Omega_{2n}+\dots+\Gamma_{n-1}\Omega_{n-1,n}.
\label{4.5}
\end{eqnarray}

Note that the singular points of (\ref{4.5}) are given with
$\Gamma_n=\dot\Gamma_n=0$. Also, equations (\ref{4.5}) have integrals
$$
F_{i,j}(\Omega)=\frac{I_i+I_n}{B_i}\Omega_{in}-\frac{I_j+I_n}{B_j}\Omega_{jn},\quad i,j=1,2,\dots,n-1.
$$

Therefore, it is natural to introduce new coordinates $\omega_1,\dots,\omega_{n-1},\gamma_1,\dots,\gamma_n$ in the
following way:
\begin{eqnarray}
&&\omega_1=\frac{I_1+I_n}{B_1} \Omega_{1n},\nonumber\\
&&\omega_i= \frac{I_i+I_n}{B_i}\Omega_{in}-\frac{I_1+I_n}{B_1}\Omega_{1n},\quad i=2,\dots,n-1,\nonumber\\
&&\gamma_1=-\frac{I_1+I_n}{B_1}\Gamma_1,\nonumber\\
&&\gamma_i=-\frac{I_i+I_n}{B_i}\Gamma_{i}+\frac{I_1+I_n}{B_1}\Gamma_{1}, \quad i=2,\dots,n-1, \nonumber \\
&& \gamma_n=\Gamma_n.
\label{4.6}
\end{eqnarray}

Then system (\ref{4.5}) get the simple form
\begin{eqnarray}
&&\frac{d\omega_1}{dt}=\gamma_n, \nonumber\\
&&\frac{d\omega_{i}}{dt}=0,\quad i=2,\dots,n-1, \nonumber\\
&&\frac{\gamma_i}{dt}=\gamma_n\omega_{i}, \quad i=1,2,\dots,n-1, \nonumber \\
&&\frac{\gamma_n}{dt}= -\left(\frac{B_1}{I_1+I_n}\right)^2\gamma_1\omega_{1}-
\sum_{i=2}^{n-1}\left(\frac{B_i}{I_i+I_n}\right)^2(\gamma_1+\gamma_i)(\omega_{1}+\omega_i).
\label{4.7}
\end{eqnarray}

Now, we can easily integrate the above system. We have $
\omega_2=\omega_2^0,\dots,\omega_{n-1}=\omega_{n-1}^0. $ By
substitution of the first equation $\dot \omega_1=\gamma_n$ to the
equations with $\gamma_i$--s we get the trajectory as a function
of the variable $\omega_1$:
\begin{eqnarray}
&&\gamma_1-\gamma_1^0=\frac12((\omega_1)^2-(\omega_1^0)^2),\nonumber\\
&&\gamma_i-\gamma_i^0=\omega_i^0(\omega_1-\omega_1^0),\quad i=2,\dots,n-1,
\label{4.8}
\end{eqnarray}
where $\gamma_1^0,\dots,\gamma_{n-1}^0$ are initial conditions.

Since $\langle\Gamma,\Gamma\rangle=1$, we have
$$
\gamma_n^2=\Gamma_n^2=1-\sum_{i=1}^{n-1}\Gamma_i^2=
1-\left(\frac{B_1}{I_1+I_n}\right)^2\gamma_1^2
-\sum_{i=2}^{n-1}\left(\frac{B_i}{I_i+I_n}\right)^2(\gamma_1+\gamma_i)^2=P_4(\omega_1),
$$
where $P_4$ is a four degree polynomial depending on initial conditions as parameters. Finally, to
get the trajectory as a function of the time $t$ one should solve
the elliptic integral
$$
t-t_0=\int_{\omega_1^0}^{\omega_1} \frac{dw_1}{\pm\sqrt{P_4(\omega_1)}}.
$$

The curve $(\gamma_1(\omega_1),\dots,\gamma_{n-1}(\omega_1))$ lies within the ellipsoid
$$
\Sigma=\left\{\left(\frac{B_1}{I_1+I_n}\right)^2\gamma_1^2+
\sum_{i=2}^{n-1}\left(\frac{B_i}{I_i+I_n}\right)^2(\gamma_1+\gamma_i)^2=1\right\}.
$$

On $\Sigma$ we have $\gamma_n=\Gamma_n=0$. Therefore the variable
$\omega_1$ varies in the
interval $[\xi_1,\xi_2]$ where  $\xi_1$ and $\xi_2$ are adjacent roots of  the polynomial $P_4$
between which it takes positive values and  $(\gamma_1(\xi_i),\dots,\gamma_{n-1}(\xi_i))\in \Sigma$. Thus,
the trajectory is either asymptotic or periodic with period
$$
T=2\int_{\xi_1}^{\xi_2}\frac{d\omega_1}{\sqrt{P_4(\omega_1)}}.
$$

We can summarize the above considerations in the following theorem

\begin{theorem}
The phase space (\ref{4.1a}) of the  multidimensional Kharlamova system is
almost everywhere foliated by
invariant circles. The periodic motions can be expressed as elliptic functions of time $t$.
\end{theorem}

\subsection{The Klebsh--Tisserand case}

By multidimensional Klebsh--Tisserand case we mean the nonholonomic rigid
body motion (\ref{4.3})
in the presence of the quadratic potential  $V(\Gamma)=\frac12(B_1\Gamma_1^2+\dots+B_n\Gamma_n^2)$:
\begin{eqnarray}
&&(I_i+I_n)\dot\Omega_{in}=(B_1-B_n)\Gamma_i\Gamma_n, \nonumber\\
&&\dot \Gamma_i=- \Gamma_n\Omega_{in}, \quad i=1,2,\dots,n-1, \nonumber \\
&& \dot \Gamma_n= \Gamma_1\Omega_{1n}+\Gamma_2\Omega_{2n}+ \dots+\Gamma_{n-1}\Omega_{n-1,n}.
\label{4.9}
\end{eqnarray}

The system (\ref{4.9}) is similar to the  Fedorov--Kozlov integrable case \cite{FK}. We have integrals
$$
F_{i}(\Omega,\Gamma)=(B_i-B_n)\Gamma_i^2+(I_i+I_n)\Omega_{in}^2 \quad i=1,2,\dots,n-1.
$$
Note that the Hamiltonian function can be expressed in terms of $F_i$--s:
$E=\frac12(nB_n+F_1+\dots+F_{n-1})$.

The phase space (\ref{4.1a}) of multidimensional  Klebsh-Tisserand system is foliated by invariant varieties
$$
\mathcal T_c=\{F_1=c_1,\dots,F_{n-1}=c_n\}.
$$

Suppose that $B_i>B_n$, $i=1,\dots,n-1$. Then the functions $F_i$ are positive and
$S_{c_i}^1=\{F_i=c_i\}$ are circles in the planes $\R^2\{\Omega_{in},\Gamma_i\}$, $i=1,2,\dots,n-1$.
The invariant surface  $\mathcal T_c$ is a two--covering of
$$
S^1_{c_1}\times\dots\times S^1_{c_{n-1}}\cap \{\Gamma_1^2+\dots+\Gamma_{n-1}^2\le 1\}
$$
determined by the function $\Gamma_n=\pm \sqrt{1-(\Gamma_1^2+\dots+\Gamma_{n-1}^2)}$. The branching
points of the covering correspond to the zeros of $\Gamma_n$.
In particular, if constants $c_i$ satisfy inequality
\begin{equation}
{\frac{c_1}{B_i-B_n}}+\dots+{\frac{c_{n-1}}{B_{n-1}-B_n}} <1,
\label{4.10}
\end{equation}
then $\Gamma_n\ne 0$ and $\mathcal T_c$ is the union of two disjoint ($n-1$)--dimensional tori
(or less dimensional tori if some of $c_i$--s are equal to zero).

The motion on $\mathcal T_c$ can be solved in term of the new time $d\tau=\Gamma_n dt$.
Define the angular variables $\varphi_i$ on circles $S^1_{c_i}$ by putting
$$
\Omega_{in}=\sqrt{\frac{c_i}{I_i+I_n}}\sin\varphi_i,
\quad \Gamma_{i}=\sqrt{\frac{c_i}{B_i-B_n}}\cos\varphi_i.
$$
Using (\ref{4.9}), we obtain the following quasi-periodic dynamics on $\mathcal T_c$:
$$
\frac{\varphi_i}{d\tau}=\omega_i=\sqrt{\frac{B_i-B_n}{I_i+I_n}}, \quad i=1,2,\dots,n-1.
$$

The frequencies $\omega_i$ depend only of $I_i$ and $B_i$. If the trajectories are periodic on one
torus, they are periodic on the rest of the tori as well.
Also, in the original time, system (\ref{4.9}) takes the form (\ref{d1.4})
$\dot \varphi_i=\omega_i/\Gamma_n^{-1}$.

\begin{theorem} If $c_1,\dots,c_{n-1}$ satisfy the inequality (\ref{4.10}) then the invariant variety
$\mathcal T_c$ is the  disjoint union of two invariant tori. The multidimensional Klebsh--Tisserand
system (\ref{4.9}) on $\mathcal T_c$ is completely integrable.
\end{theorem}

\subsection*{Acknowledgments}
I would like to thank
M. Radnovi\' c for the kind support.
The research was partially supported by the Serbian Ministry of Science and Technology,
Project 1643 -- Geometry and Topology of Manifolds and Integrable Dynamical Systems.

\end{document}